\documentstyle[preprint,eqsecnum,aps]{revtex}
\tightenlines
\draft

\newcommand{\be}{\begin{equation}}
\newcommand{\ee}{\end{equation}}
\newcommand{\beal}{\begin{eqalign}}
\newcommand{\eeal}{\end{eqalign}}
\newcommand{\bea}{\begin{eqnarray}}
\newcommand{\eea}{\end{eqnarray}}
\newcommand{\bean}{\begin{eqnarray*}}
\newcommand{\eean}{\end{eqnarray*}}
\newcommand{\ba}{\begin{array}}
\newcommand{\ea}{\end{array}}

\newcommand{\ep}{\epsilon}

\newcommand{\ka}{\theta}

\newcommand{\La}{\Lambda}

\newcommand{\de}{\delta}

\newcommand{\pa}{\partial}

\newcommand{\no}{\nonumber}

\newcommand{\tr}{\mbox{tr}}
\newcommand{\res}{\mbox{res}}

\begin{document}

\title
 { \sc Conformal Covariantization of Moyal-Lax Operators\/}
\author{
{\sc Ming-Hsien Tu$^\dag$
\footnote{E-mail: phymhtu@ccunix.ccu.edu.tw (Corresponding author)}\/}
, Niann-Chern Lee$^\ddag$ and Yu-Tung Chen$^\dag$\\
  {\it $^\dag$Department of Physics, National Chung Cheng University,\\
   Minghsiung, Chiayi 621, Taiwan\/}\\
   {\it $^\ddag$General Education Center, National Chinyi Institute of Technology,\\
    Taiping, Taichung 411, Taiwan\/}\\
   }
\date{\today}
\maketitle
\begin{abstract}
A covariant approach to the conformal property associated with Moyal-Lax operators
is given. By identifying the conformal covariance with the second Gelfand-Dickey
flow, we covariantize  Moyal-Lax operators to construct the primary fields
of one-parameter deformation of classical $W$-algebras.
   \end{abstract}

\newpage
\section{introduction}
Recently, there has been a great deal of interest to study the Moyal deformation
of the KdV equations in variant ways, such as Lax and/or Hamiltonian formulations
\cite{Ku,St,Tu,DP,DP2}, zero-curvature formulation \cite{Ko}, B\"acklund transformation
 \cite{Tu2} and classical Virasoro and $W$-algebras \cite{Tu,DP}, etc.
In these formulations, the ordinary (pseudo-) differential Lax operator
$L=\sum_iu_i(x)\pa^i$ is replaced by (pseudo-) differential symbols $M(x,p)$, the
formal Laurent series in $p$, which obey a noncommutative but associative algebra
 with respect to the $\star$-product \cite{Gr} defined by
\be
M(x,p)\star N(x,p)=\sum_{s=0}^{\infty}\frac{\ka^s}{s!}\sum_{j=0}^s(-1)^j{s \choose j}
(\pa_x^j\pa_p^{s-j}M)(\pa_x^{s-j}\pa_p^jN),
 \label{sp}
\ee
where $\ka$ is a dimensionless parameter characterizing the strength of the deformation.
On the other hand, by (\ref{sp}), the ordinary commutator is thus taken over by the Moyal
 bracket \cite{M}
\be
 \{M(x,p),N(x,p)\}_\ka=\frac{M\star N-N\star M}{2\ka},
 \label{Moyalb}
\ee
that possesses the anti-symmetry, bi-linearity and Jacobi identity.
 The  Moyal bracket (\ref{Moyalb}) can be viewed as the higher-order derivative
  (or dispersive) generalization of the canonical
  Poisson bracket since it recovers the canonical Poisson  bracket in the limit $\ka\to 0$,
  namely, $\lim_{\ka \to 0}\{M,N\}_\ka=\pa_pM\pa_xN-\pa_xM\pa_pN$.
  It turns out that the Moyal formulation of Lax equations reduces to dispersionless
   Lax equations \cite{Z,KG,Kr,FR,TT} in this limit.

   In the previous work \cite{Tu} we have studied the $W$-algebraic structure associated with the
   Moyal formulation of the KdV equations. We worked out the Poisson brackets of the
   second Gelfand-Dickey (GD) structure \cite{GD,D} defined by the $\star$-product
   and obtained an one-parameter deformation of
   the classical $W_n$-algebra including a Virasoro subalgebra with central charge
   $\ka^2(n^3-n)/3$. In this work, we would like to investigate further the $W$-algebraic
    structure from the point of view of conformal covariance. We shall follow the approach
    developed by Di Francesco, Itzykson and Zuber (DIZ) \cite{DIZ}
    to covariantize the Moyal-Lax operators \cite{Ku,DP} and identify the underlying
    primary fields in a systematic way.

This paper is organized as follows. In section II, we recall the Moyal-Lax
formulation of the KdV equations using psuedo-differential symbols with
 respect to the $\star$-product. We then introduce
the second GD structure defined by the Moyal bracket and show that it indeed provides
the Hamiltonian structure for the Moyal-type Lax equations. In section III, the
diffeomorphism$(S^1)$ is defined and the conformal transformation of the Moyal-Lax
 operators is investigated.
Then in section IV, we show that the infinitesimal diffeomorphism flow defined by
conformal covariance is equivalent to that of the Hamiltonian flow defined by the second
 GD structure. This enables us to define the primary fields of the diffeomorphism.
 Following DIZ, in section V, we systematically covariantize the Moyal-Lax operator to
  decompose the coefficient functions of the Lax operator into the conformal primary fields
   which satisfy an one-parameter deformation of the classical $W_n$-algebra including
   a Virasoro subalgebra. In section VI the covariantization is generalized to
   psuedo-differential symbols to construct more additional primary fields. Section
VII is devoted to the conclusions and discussions.
\section{Lax equations and Hamiltonian structures}
For the differential symbol $L=p^n+\sum_{i=1}^{n}u_i\star p^{n-i}$ with coefficients
$u_i$ depending on an infinite set of variables $x\equiv t_1, t_2, t_3,\cdots$ one can define
the Lax equations \cite{Ku,DP}
 \bea
  \frac{\pa L}{\pa t_k}&=&\{(L^{1/n}\star)^k_+,L\}_{\ka},\qquad
 (L^{1/n}\star)^k_+=(\underbrace{L^{1/n}\star L^{1/n}\star\cdots \star L^{1/n}}_{k})_+,\no\\
 &=&\{L,(L^{1/n}\star)^k_-\}_\ka,
 \label{laxeq}
\eea
where  $L^{1/n}=p+\sum_{i=0}a_i\star p^{-i}$ is   the $n$th root of $L$ in such a way that $
L=(L^{1/n}\star)^n$ and $(A)_{+/-}$ refer to the non-negative/negative powers in $p$
of the pseudo-differential symbol $A$. Note that the evolution equation for $u_1$
is trivial since the highest order in $p$
 on the right-hand side of the Lax equations (\ref{laxeq}) is $n-2$ due to the
 definition of the Moyal bracket, and hence one can drop $u_1$ in the Lax formulation.
 However, we will see that this is not the case for the Hamiltonian formulation.

Next let us formulate the Lax equations (\ref{laxeq}) in terms of Hamiltonian structure.
 For the functionals $F[L]$ and $G[L]$ we define the second Gelfand-Dickey
   bracket \cite{D} with respect to the  $\star$-product as
  \be
\{F,G\}_2=\tr[J^{(2)}(d_LF)\star d_LG]=\int \res[J^{(2)}(d_LF)\star d_LG],
 \label{gdb}
  \ee
 where $\res (A)=a_{-1}$ and $\tr(A)=\int \res(A)$ denote the residue and trace of
  $A=\sum_i a_i\star p^i$, and $J^{(2)}$ is the Adler map \cite{A} defined by
   \bea
  J^{(2)}(d_LF)&=&\{L,d_LF\}_{\ka+}\star L-\{L,(d_LF\star L)_+\}_\ka,\no\\
  &=&\{L,(d_LF\star L)_-\}_\ka-\{L,d_LF\}_{\ka-}\star L,
  \label{adler}
  \eea
  with $d_LF\equiv \de F/\de L=\sum_{i=1}^np^{-n+i-1}\star\de F/\de u_i$. The bracket defined by
 $J^{(2)}$ is indeed Hamiltonian since $\{F,G\}_2=-\{G,F\}_2$ due to the cyclic property of
  the trace and  the Jacobi identity can be verified \cite{Tu2} by
   the Kupershmidt-Wilson (KW) theorem \cite{KW}. Form (\ref{adler}) $J^{(2)}(X)$ is
    linear in $X$ and has order at most $n-1$. One can use the standard Dirac procedure
     \cite{DIZ} to get rid of $u_1$ so that
 \be
 \hat{J}^{(2)}(X)=\{L,X\}_{\ka+}\star L-\{L,(X\star L)_+\}_\ka+
 \frac{1}{n}\{L,\int^x\res\{L,X\}_\ka\}_\ka
 \label{rgd}
 \ee
 or, in components, $\hat{J}^{(2)}(X)=\sum_{i,j=2}^{n}(\hat{J}^{(2)}_{ij}\cdot x_j)\star p^{n-i}$
where $\hat{J}^{(2)}_{ij}$ are differential operators, and
hence the reduced Poisson brackets for $u_i$ can  be expressed as $
\{u_i(x),u_j(y)\}^D_2=\hat{J}^{(2)}_{ij}\cdot\delta(x-y)$.
From the reduced GD brackets (\ref{rgd}) the Hamiltonian flows can be expressed as
 \be
\frac{\pa L}{\pa t_k}=\{L,H_k\}_2^D=\hat{J}^{(2)}(d_LH_k),
\label{hameq}
 \ee
where the Hamiltonians $H_k$ are defined by
 \be
  H_k=\frac{n}{k}\int \res(L^{1/n}\star)^k.
\label{ham}
 \ee
Using (\ref{ham}) and the fact $d_LH_k=(L^{1/n}\star)^{k-n}_-$ mod $p^{-n}$ it
is straightforward to show that the Hamiltonian flows (\ref{hameq})
are equivalent to the Lax equations (\ref{laxeq}).
\section{Diff$(S^1)$ and conformal covariance}
A function $f(x)$ is a primary field with conformal weight $h$ if under the diffeomorphism
$x\to t(x)$ it transforms as
\be
f(x)\to \tilde{f}(t)=\phi^{-h}f(x)=\phi^{-h}\star f(x),
\label{primary}
\ee
where $\phi(x)\equiv dt(x)/dx$. We denote ${\mathcal F}_h$ the space of functions
 with weight $h$ (or spin-$h$ primary fields).
 For a covariant operator $\Delta(x,p)$ that maps ${\mathcal F}_h$
  to ${\mathcal F}_l$, then it transforms according to
\be
\tilde{\Delta}(t,p_t)=\phi^{-l}\star\Delta(x,p)\star\phi^h,
\label{operator}
\ee
where $p_t=\phi^{-1}\star p$ is the conjugate momentum of $t$ with respect to the Moyal bracket,
i.e. $\{p_t,t\}_\ka=1$ and has an inverse $p_t^{-1}=p^{-1}\star \phi$.

Let us treat the Lax operator $L_n(x,p)=p^n+u_2(x)\star p^{n-2}+\cdots+u_n(x)$ as
a covariant operator such that $L_n(x,p): {\mathcal F}_h \to {\mathcal F}_l$.
The corresponding weights $h$ and $l$ have to be determined from the transformation law:
\bea
\tilde{L}_n(t,p_t)&=&\phi^{-l}\star L_n(x)\star \phi^h,\no\\
&=&(p_t\star)^n+\tilde{u}_2(t)\star(p_t\star)^{n-2}+\cdots+\tilde{u}_n(t).
\label{cov-lax}
\eea
We note that $p_t=\phi^{-1}\star p=(\sqrt{\phi})^{-1}\star \phi^{-1}p\star \sqrt{\phi}$
which, by induction, gives
\be
(p_t\star)^k=\frac{1}{\sqrt{\phi}}\star [\phi^{-k}p^k+\frac{\ka^2f_k}{\phi^k}p^{k-2}+\cdots]
\star \sqrt{\phi},
\label{dress}
\ee
with
\[
f_k=-\frac{k(k-1)}{2}\left(\frac{\phi'}{\phi}\right)^2-\frac{k(k-1)(k-2)}{6}\frac{\phi''}{\phi}.
\]
Substituting (\ref{dress}) into (\ref{cov-lax}) we have
$h=-(n-1)/2,\quad l=(n+1)/2$
and $u_2(x)$ transforms like an anomalous spin-2 primary field
\be
\tilde{u}_2(t)=\phi^{-2}u_2(x)+\frac{\ka^2(n^3-n)}{3}\{\{x,t(x)\}\},
\label{conf-u2}
\ee
where $\{\{x,t(x)\}\}$ is the schwarzian derivative defined by
\be
\{\{x,t(x)\}\}=\left(\frac{\frac{d^3x}{dt^3}}{\frac{dx}{dt}}\right)-
\left(\frac{\frac{d^2x}{dt^2}}{\frac{dx}{dt}}\right)^2
=\frac{\phi''}{\phi^3}-\frac{3}{2}\left(\frac{\phi'}{\phi^2}\right)^2.
\label{schwarz}
\ee
Eq.(\ref{conf-u2}) indicates that $u_2$ can be viewed as the generator of the classical
 Virasoro algebra with central charge $c_{n,\ka}=\ka^2(n^3-n)/3$.

\section{Virasoro flows as Hamiltonian flows}

As we shown in the previous section that it is quite difficult to obtain the
 transformation laws for $u_{i>2}$ under the finite diffeomorphism.  However it is
 manageable to investigate the infinitesimal transformations of $u_i$. For an
 infinitesimal diffeomorphism $x\to t(x)\simeq x-\ep(x)$ we have
  $\phi(x)\simeq 1-\ep'(x)$ and $p_t=p+\{p,\ep\}_\ka\star p$. In particular,
  it can be easily proved by induction that $(p_t\star)^i=p^i+\{p^i,\ep\}_\ka\star p$. Hence
from (\ref{cov-lax}) we have
\bean
\tilde{L}_n(t)&=&\sum_i(u_i(x)-\ep(x)u_i'(x)+\de_\ep u_i(x))\star(p^i+\{p^i,\ep\}_\ka\star p),\\
&=&L_n(x)+\{L_n(x),\ep(x)\}_\ka\star p-\ep(x)\star L_n'(x)+\de_\ep L_n(x),\\
&=&\left(1+\frac{n+1}{2}\ep'(x)\right)\star L_n(x)\star\left(1+\frac{n-1}{2}\ep'(x)\right),\\
&=&L_n(x)+\frac{n+1}{2}\ep'(x)\star L_n(x)+\frac{n-1}{2}L_n(x)\star\ep'(x),
\eean
which leads to the infinitesimal change of the Lax operator
\bea
\de_\ep L_n(x)&=&\frac{n+1}{2}\ep'(x)\star L_n(x)+\frac{n-1}{2}L_n(x)\star\ep'(x)\no\\
&&-\{L_n(x),\ep(x)\}_\ka\star p+\ep(x)\star\{p,L_n(x)\}_\ka.
\label{inf-conf}
\eea
Next let us consider the Hamiltonian flow generated by the Hamiltonian $H=\int \ep(x)u_2(x)dx$.
From the second GD structure (\ref{rgd}) and Hamiltonian flow (\ref{hameq}) we have
\bean
\de^{GD}L_n(x)&=&\{L_n(x),X\}_\ka\star L_n(x)-\{L_n(x),(X\star L_n(x))_+\}_\ka\no\\
&&+\frac{1}{n}\{L_n(x), \int^x\res\{L_n(x), X\}_\ka\}_\ka,
\eean
where $X\equiv \de H/\de L=p^{-n+1}\star\ep(x)$.
A simple algebra shows that
\bean
&&(L_n\star X)_+=p\star\ep,\\
&&(X\star L_n)_+=\ep\star p-2\ka(n-1)\ep',\\
&&\frac{1}{n}\{L_n, \int^x\res\{L_n, X\}_\ka\}_\ka=
-\frac{n-1}{2}(L_n\star\ep'-\ep'\star L_n),
\eean
which implies
\bean
\de^{GD}L_n&=&\frac{1}{2\ka}[p\star\ep\star L_n-
L_n\star(\ep\star p-2\ka(n-1)\ep')]-\frac{n-1}{2}(L_n\star\ep'-\ep'\star L_n),\\
&=&\de_\ep L_n
\eean
as desired. Comparing the both hand sides of (\ref{inf-conf}) we get the infinitesimal variations
of $u_k$ $(2\leq k\leq n)$ as
\bea
\de_\ep u_k&=&u_k'\ep+ku_k\ep'+\frac{(2\ka)^k(k-1)}{2}{n+1\choose k+1}\ep^{(k+1)}\no\\
&&+\sum_{i=2}^{k-1}(2\ka)^{k-i}\left[\frac{n-1}{2}{n-i\choose k-i}-{n-i\choose k-i+1}\right]
u_i\ep^{(k-i+1)},
\label{inf-var}
\eea
where ${n\choose m}$  are the standard binomial coefficients with $0\leq m\leq n$.
 Let us list  the first few $\de u_k$:
\bea
\de_\ep u_2&=&u_2'\ep+2u_2\ep'+\frac{\ka^2(n^3-n)}{3}\ep''',\no\\
\de_\ep u_3&=&u_3'\ep+3u_3\ep'+2\ka(n-2) u_2\ep''+\frac{\ka^3(n^3-n)(n-2)}{3}\ep^{(4)},\no\\
\de_\ep u_4&=&u_4'\ep+4u_4\ep'+3\ka(n-3) u_3\ep''+\frac{\ka^2(n-2)(n-3)(n+5)}{3}u_2\ep'''\no\\
&&+\frac{\ka^4(n-2)(n-3)(n^3-n)}{5}\ep^{(5)},\no\\
\de_\ep u_5&=&u_5'\ep+5u_5\ep'+4\ka(n-4) u_4\ep''+\frac{\ka^2(n-3)(n-4)(n+7)}{3}u_3\ep'''\no\\
&&+\frac{\ka^3(n-2)(n-3)(n-4)(n+3)}{3}u_2\ep^{(4)}+
\frac{4\ka^5(n-2)(n-3)(n-4)(n^3-n)}{45}\ep^{(6)},
\label{inf-u}
\eea
etc. The first equation in (\ref{inf-u}) is just the infinitesimal version of (\ref{conf-u2})
 which together with the  Hamiltonian flow $
\de_\ep u_2(x)=\{u_2(x), H\}_2^D=\int\{u_2(x),u_2(y)\}_2^D\ep(y)dy$
 implies the classical Virasoro algebra
\be
\{u_2(x),u_2(y)\}_2^D=[c_{n,\ka}\pa_x^3+2u_2\pa_x+u_2']\de(x-y).
\label{virasoro}
\ee
Furthermore,  it has a simple interpretation about  the other relations in (\ref{inf-u}).
We can define a new variable $w_k=u_k+f(u_i)$, where $f(u_i)$ is a differential
polynomial in $u_{i< k}$, such that $w_k$ is a spin-$k$ primary field with
respect to the generator $u_2$, namely,
\[
\{w_k(x),u_2(y)\}_2=[kw_k\pa_x+w_k']\de(x-y).
\]
 For instance, let $w_3=u_3+\alpha u_2'$ and demanding the
relation $\de_\ep w_3=\ep w_3'+3w_3\ep'$ then we get $\alpha=-\ka(n-2)$. On the other hand,
let $w_4=u_4+\alpha u_3'+\beta u_2''+\gamma u_2^2$ and demanding the relation
$\de_\ep w_4=\ep w_4'+4w_4\ep'$ we have $\alpha=-\ka(n-3)$, $\beta=2\ka^2(n-2)(n-3)/5$
 and $\gamma=-(n-2)(n-3)(5n+7)/[10(n^3-n)]$.

In summary, we can identify the following primary fields
\bea
w_3&=&u_3-\ka(n-2)u_2',\no\\
w_4&=&u_4-\frac{(n-2)(n-3)(5n+7)}{10(n^3-n)}u_2^2-\ka(n-3)u_3'+\frac{2\ka^2(n-2)(n-3)}{5}u_2'',\no\\
w_5&=&u_5-\ka(n-4)u_4'+\frac{3\ka^2(n-3)(n-4)}{7}u_3''-\frac{2\ka^3(n-2)(n-3)(n-4)}{21}u_2'''\no\\
&&+\frac{(n-3)(n-4)(7n+13)}{7(n^3-n)}[\ka(n-2)u_2u_2'-u_2u_3],
\label{def-w}
\eea
etc.
To construct the primary fileds $w_k$ for $k>5$ we shall
covariantize the Lax operator in a systematic way.

\section{Covariantizing the Lax operators}

For a series of change of variable $v\to x\to t$, the schwarzian derivative obeys
 the equation
 \be
\{\{v,t\}\}=\left(\frac{dx}{dt}\right)^2\{\{v,x\}\}+\{\{x,t\}\},
\label{conf-schwarz}
 \ee
which, comparing with (\ref{conf-u2}), shows that $u_2(x)$ transforms as $c_{n,\ka}\{\{v,x\}\}$.
Define the variable $b(x)=\frac{d^2v}{dx^2}(\frac{dv}{dx})^{-1}$ it turns out that,
 for $n\neq -1,0, 1$ and $\ka\neq 0$
\be
\frac{u_2(x)}{c_{n,\ka}}=\{\{v,x\}\}=b'(x)-\frac{1}{2}b^2(x),
\label{def-b}
\ee
with $v$ being the coordinate where $u_2$ vanishes, i.e. $u_2(v)=0$.
 It is easy to show that $b(x)$ transforms as an anomalous spin-1 primary field
\be
\tilde{b}(t)=\frac{d^2v}{dt^2}\left(\frac{dv}{dt}\right)^{-1}=
\frac{dx}{dt}b(x)+\frac{d^2x}{dt^2}\left(\frac{dx}{dt}\right)^{-1}.
\label{tran-b}
\ee
The purpose for introducing $b(x)$ is to construct a covariant operator
$D_k=p-2\ka kb(x)$ which maps ${\mathcal F}_k$ to ${\mathcal F}_{k+1}$. Using
$D_k$ the covariant operator $D_k^l:{\mathcal F}_k \to {\mathcal F}_{k+l}$
can be constructed as $D_k^l=D_{k+l-1}\star D_{k+l-2}\star\cdots\star D_k (l>1)$.

Now, following DIZ procedure, the Lax operator $L_n$  can be decomposed
into the sum of the covariant operators $\Delta_k^{(n)}:{\mathcal F}_{-\frac{n-1}{2}} \to
 {\mathcal F}_{\frac{n+1}{2}}$ as
 \be
L_n=\Delta_2^{(n)}(u_2)+\Delta_3^{(n)}(w_3,u_2)+\cdots +\Delta_n^{(n)}(w_n,u_2),
\label{decomp-lax}
 \ee
where
\bean
\Delta_2^{(n)}&=&D_{-\frac{n-1}{2}}^{n}=[p-\ka(n-1)b(x)]\star[p-\ka(n-3)b(x)]
\star\cdots\star[p+\ka(n-1)b(x)],\\
\Delta_k^{(n)}&=&\sum_{l=0}^{n-k}\alpha_{k,l}^{(n)}(D_k^l\star w_k)\star
D_{-\frac{n-1}{2}}^{n-k-l},
\eean
and the coefficients $\alpha_{k,l}^{(n)}$ are determined from the requirement that
the Lax operator $L_n$ depends on $u_2$ only through the relation (\ref{def-b}).
 Therefore the function $b(x)$ is defined up to the condition $(\de b)'-b\de b=0$ or
  equivalently, $D_{k+1}\star \de b=\de b\star D_k$. In particular we have
\bea
\de_b D_k^l&=&\sum_{i=1}^l D_{k+l-1}\star\cdots\star\de_bD_{k+l-i}\star\cdots\star D_k,\no\\
&=&\sum_{i=1}^l D_{k+l-1}\star\cdots\star[-2\ka(k+l-i)\de b]\star\cdots\star D_k,\no\\
&=&-\ka l(2k+l-1)\de b\star D_k^{l-1}.
\label{var-D}
\eea
Hence $\de_b L_n=0$ implies
\be
\de_bD_{-\frac{n-1}{2}}^n+\sum_{k=3}^n\sum_{l=0}^{n-k}\left[\alpha_{k,l}^{(n)}
(\de _bD_k^l\star w_k)\star D_{-\frac{n-1}{2}}^{n-k-l}+
\alpha_{k,l}^{(n)}(D_k^l\star w_k)\star \de_b D_{-\frac{n-1}{2}}^{n-k-l}\right]=0.
\label{var-lax}
\ee
From (\ref{var-D}) it is easy to show that the first term in (\ref{var-lax}) vanishes.
 For those terms in summation we get the recursive relation
\[
\alpha_{k,l+1}^{(n)}=\frac{(k+l)(n-k-l)}{(2k+l)(l+1)}\alpha_{k,l}^{(n)},\qquad k\geq 3
\]
which together with the normalization condition $\alpha_{k,0}^{(n)}=1$ yields
\[
\alpha_{k,l}^{(n)}=\frac{{k+l-1\choose l}{n-k \choose l} }{{ 2k+l-1\choose l}}.
\]
Let us work out the first few terms for the decomposition (\ref{decomp-lax}).
 A straightforward computation yields
\bea
(D_k\star w_k)&=&2\ka(w_k'-kbw_k),\no\\
(D_k^2\star w_k)&=&4\ka^2[w_k''-(2k+1)bw_k'+(k(k+1)b^2-kb')w_k],\no\\
(D_k^3\star w_k)&=&8\ka^3[w_k'''-3(k+1)bw_k''-(3k+1)b'w_k'+(3k^2+6k+2)b^2w_k'\no\\
&&-k(k+1)(k+2)b^3w_k+k(3k+4)bb'w_k-kb''w_k],
\label{identity-1}
\eea
and
\bean
D_{-\frac{n-1}{2}}^n&=&p^n+u_2\star p^{n-2}+\ka (n-2)u_2'\star p^{n-3}\\
&&+\left[\frac{3\ka^2(n-2)(n-3)}{5}u_2''+\frac{(n-2)(n-3)(5n+7)}{10(n^3-n)}u_2^2\right]
\star p^{n-4}\\
&&+\left[\frac{\ka(n-2)(n-3)(n-4)(5n+7)}{5(n^3-n)}u_2u_2'+
\frac{4\ka^3(n-2)(n-3)(n-4)}{15}u_2'''\right]\star p^{n-5}+\cdots,\\
D_{-\frac{n-1}{2}}^{n-3}&=&p^{n-3}+3\ka(n-3)b \star p^{n-4}\\
&&+\left[\frac{\ka^2(n-3)(n-4)(n+7)}{3}b'-\frac{(n-3)(n-4)(n-29)}{6}b^2\right]
\star p^{n-5}+\cdots,\\
D_{-\frac{n-1}{2}}^{n-4}&=&p^{n-4}+4\ka(n-4)b\star p^{n-5}+\cdots,\\
D_{-\frac{n-1}{2}}^{n-5}&=&p^{n-5}+\cdots.
\eean
Thus
\bean
\Delta_2^{(n)}(u_2)&=&D_{-\frac{n-1}{2}}^n,\\
\Delta_3^{(n)}(w_3,u_2)&=&w_3\star p^{n-3}+\ka(n-3)w_3'\star p^{n-4}\\
&&+\left[\frac{4\ka^2(n-3)(n-4)}{7}w_3''+
\frac{(n-3)(n-4)(7n+13)}{7(n^3-n)}u_2w_3\right]\star p^{n-5}+\cdots,\\
\Delta_4^{(n)}(w_4,u_2)&=&w_4\star p^{n-4}+\ka(n-4)w_4'\star p^{n-5}+\cdots,\\
\Delta_5^{(n)}(w_5,u_2)&=&w_5\star p^{n-5}+\cdots,
\eean
which decomposes the coefficient functions $u_i$ into the primary fields
\bea
u_2&=&w_2,\no\\
u_3&=&w_3+\ka(n-2)u_2',\no\\
u_4&=&w_4+\ka(n-3)w_3'+\frac{3\ka^2(n-2)(n-3)}{5}u_2''+\frac{(n-2)(n-3)(5n+7)}{10(n^3-n)}u_2^2,\no\\
u_5&=&w_5+\ka(n-4)w_4'+\frac{4\ka^2(n-3)(n-4)}{7}w_3''+\frac{(n-3)(n-4)(7n+13)}{7(n^3-n)}w_3u_2\no\\
&&+\frac{\ka(n-2)(n-3)(n-4)(5n+7)}{5(n^3-n)}u_2u_2'+\frac{4\ka^3(n-2)(n-3)(n-4)}{15}u_2'''.
\label{decomp-u}
\eea
Inverting the above relation we recover the definition (\ref{def-w}) of the primary fields.

\section{generalizations}
In this section we would like to show that the conformal covariantization for the Lax
operator (\ref{cov-lax}) can be extended to a more general form
\be
\La_n=p^n+u_2\star p^{n-2}+\cdots+u_n+u_{n+1}\star p^{-1}+u_{n+2}\star p^{-2}+\cdots.
\label{pdo-lax}
\ee
It is not hard to show that, for the pseudo-differential symbol (\ref{pdo-lax}), the
associated Hamiltonian structure is defined by the reduced Adler map (\ref{rgd}) as well.
Due to the fact that  $(\La_n)_+$ and $(\La_n)_-$ are transformed independently under
 (\ref{cov-lax}), the infinitesimal change of $u_k (2\leq k\leq n )$ is the same as
 (\ref{inf-u}), while that of  $u_k (k\geq n+1 )$, governed by (\ref{inf-conf}), yields
  \[
\de u_{n+k}=u_{n+k}'\ep+(n+k)u_{n+k}\ep'+\sum_{i=1}^{k-1}(2\ka)^{k-i}
\left[\frac{n-1}{2}{-i \choose k-i}-{-i \choose k-i+1}\right]u_{n+i}\ep^{(k-i+1)},
  \]
 where ${-n\choose m}\equiv (-1)^m{n+m-1\choose m}$ with $n,m\geq 0$; from which,
  the following primary fields can be defined
\bea
w_{n+1}&=&u_{n+1},\no\\
w_{n+2}&=&u_{n+2}+\ka u_{n+1}',\no\\
w_{n+3}&=&u_{n+3}+2\ka u_{n+2}'+\frac{2\ka^2(n+1)}{2n+3}u_{n+1}''-
\frac{6(n+1)}{n(n-1)(2n+3)}u_2u_{n+1},\no\\
w_{n+4}&=&u_{n+4}+3\ka u_{n+3}'+\frac{6\ka^2(n+2)}{2n+5}u_{n+2}''
+\frac{2\ka^3(n+1)}{2n+5}u_{n+1}'''\no\\
&&-\frac{6(3n+7)}{n(n-1)(2n+5)}u_2u_{n+2}
-\frac{6(3n+7)}{n(n-1)(2n+5)}u_2u_{n+1}',
\label{def-wn}
\eea
etc. To covariantize the negative part $(\La_n)_-$ one can define the covariant operator
 $D_k^{-1}:{\mathcal F\/}_k\to {\mathcal F\/}_{k-1}$ as \cite{H}
 \be
D_k^{-1}\equiv [D_{k-1}]^{-1}=p^{-1}+2\ka(k-1)b\star p^{-2}+\cdots,
\label{def-Dn}
\ee
and thus
\be
D_k^{-l}=[D_{k-l}^l]^{-1}=D_{k-l-1}^{-1}\star D_{k-l}^{-1}\cdots\star D_k^{-1},
\label{def-prDn}
\ee
with a covariant property determined by that of
$D_{k-l}^l$ as
\[
D_k^{-l}(t)=[D_{k-l}^l(t)]^{-1}=\phi^{l-k}\star D_{k}^{-l}(x)\star \phi^k.
\]
Now let us decompose $(\La_n)_-$ as
\be
(\La_n)_-=\sum_{l=1}^\infty \Delta_{n+k}^{(n)}(w_{n+k},u_2),
\label{cov-laxn}
\ee
where the covariant operator $\Delta_{n+k}^{(n)}(w_{n+k},u_2)$ is linear in $w_{n+k}$
and is defined by
\[
\Delta_{n+k}^{(n)}(w_{n+k},u_2)=\sum_{l=0}^\infty \beta_{n+k,l}^{(n)}
(D_{n+k}^l\star w_{n+k})\star D_{-\frac{n-1}{2}}^{-k-l},\qquad k\geq 1.
\]
The coefficients $\beta_{n+k,l}^{(n)}$ can be determined in a similar manner so that
$(\La_n)_-$ depends on $u_2$ only through (\ref{def-b}). It turns out that
\[
\beta_{n+k,l}^{(n)}=(-1)^l\frac{{k+l-1\choose l}{ n+k+l-1\choose l}}{{2n+2k+l-1\choose l}}.
\]
Following a similar procedure discussed in the previous section and
comparing (\ref{pdo-lax}) with (\ref{cov-laxn}) we get
\bea
u_{n+1}&=&w_{n+1},\no\\
u_{n+2}&=&w_{n+2}-\ka w_{n+1}',\no\\
u_{n+3}&=&w_{n+3}-2\ka w_{n+2}'+\frac{2\ka^2(n+2)}{2n+3}w_{n+1}''+
\frac{6(n+1)}{n(n-1)(2n+3)}u_2w_{n+1},\no\\
u_{n+4}&=&w_{n+4}-3\ka w_{n+3}'+\frac{6\ka^2(n+3)}{(2n+5)}w''_{n+2}+
\frac{6(3n+7)}{n(n-1)(2n+5)}u_2w_{n+2}\no\\
&&-\frac{2\ka^3(n+3)}{(2n+3)}w_{n+1}'''-\frac{18\ka(n+1)}{n(n-1)(2n+3)}
(u_2w_{n+1})'.
\eea
Inverting the above equations yield (\ref{def-wn}) as expected.
\section{conclusion and discussions}
We have discussed the covariance of the Moyal-type Lax operator under the
 diffeomorphism$(S^1)$. By comparing the infinitesimal Diff$(S^1)$-flow with
  the GD-flow we have identified the primary fields with respect to the classical
   energy-momentum generator which obeys the classical Virasoro algebra with central
   charge $c_{n,\ka}=\ka^2(n^3-n)/3$. We then follow the DIZ procedure to covariantize
   Moyal-Lax operators and identify the primary fields in a systematic way.

Few remarks are in order. First, the $w_k$ shown above  form an
 one-parameter deformation of the primary fields arising from the (pseudo-)differential
 Lax operator. In particular, the central charge $c_{n,\ka}$ can be used to characterize
 the dispersion effect since $\ka\to 0$ corresponds to the dispersionless limit of the
 Lax equation (\ref{laxeq}). Secondly, for $\ka=1/2$, the primary fields $w_k$ recover
  the standard result \cite{DIZ,H} while for $\ka=0$ (\ref{def-w}) do not directly reproduce
  to those result in  the dispersionless limit \cite{FR} in which the coefficient functions $u_k$
  are already primary fields with respect to $u_2$, the generator of the centerless
  Virasoro algebra. This is due to the fact that the parametrization (\ref{def-b}) does not work
  for $\ka=0$ and thus the associated conformal property should be traced back to the
  GD structure or infinitesimal transformation (\ref{inf-var}).
   Thirdly, in spite of covariantizing the  Lax operator $L_n=p^n+\sum_{i=2}^nu_i\star p^{n-i}$,
    the conformal property associated with the Lax operator of the form
   \be
K_n=p^n+v_2p^{n-2}+v_3p^{n-3}+\cdots+v_n
\label{Stra-lax}
   \ee
   has been investigated \cite{Tu} as well. In fact, the Lax equations defined by $K_n$
   and $L_n$ are equivalent up to the following isomorphism
   \be
v_j=\sum_{i=1}^j(-\ka)^{j-i}{n-i \choose n-j}u_i^{(j-i)},
\label{iso-KS}
   \ee
   which can be used to construct the primary fields associated with $K_n$.
For instance, from (\ref{def-w}) and (\ref{iso-KS}), the first few primary fields
can be expressed as
\bean
w_2&=&v_2,\\
w_3&=&v_3,\\
w_4&=&v_4-\frac{(n-2)(n-3)(5n+7)}{10(n^3-n)}v_2^2-\frac{\ka^2(n-2)(n-3)}{10}v_2'',
\eean
which are just those primary fields obtained in \cite{Tu}.

Finally, based on the algebra of pseudo-differential symbols with respect to the
$\star$-product, it would be intriguing to carry out the covariant approach
for reductions, truncations, and even supersymmetrization \cite{DP2} of
 the Lax operator (\ref{pdo-lax}) to construct the corresponding $W$-algebras.
 Works in these directions are now in progress.

{\bf Acknowledgments\/}\\
  M.H.T thanks the National Science Council of Taiwan
  (Grant numbers NSC 90-2112-M-194-006) for support.


\begin{thebibliography}{99}

\bibitem{Ku}
B.A. Kupershmidt,  Lett. Math. Phys. 20 (1990) 19.

\bibitem{St}
I.A.B. Strachan, J. Phys. A 28 (1995) 1967.

\bibitem{Tu}
M.H. Tu, Phys. Lett. B 508 (2001) 173.

\bibitem{DP}
A. Das, Z. Popowicz, Phys. Lett. B 510 (2001) 264.

\bibitem{DP2}
A. Das, Z. Popowicz, J. Phys. A 34 (2001) 6105.

\bibitem{Ko}
T. Koikawa, Progr. Theoret. Phys. 105 (2001) 1045.

\bibitem{Tu2}
M.H. Tu, J. Phys. A 34 (2001) L623.

\bibitem{Gr}
H. Groenewold, Physica, 12 (1946) 405.

\bibitem{M}
J.E. Moyal, Proc. Cambridge Phil. Soc. 45 (1949) 90.

\bibitem{Z}
V.E. Zakharov, Func. Anal. Appl. 14 (1980) 89.

\bibitem{KG}
Y. Kodama, J. Gibbons, in ``Proceedings of Workshop on Nonlinear and Turbulent
Processes in Physics", World Scientific, Singapore, 1990.

\bibitem{Kr}
I. Krichever, Commun. Math. Phys. 143 (1992) 415.

\bibitem{FR}
J.M. Figueroa-O'Farrill, E. Ramos, Phys. Lett. B 282 (1992) 357.

\bibitem{TT}
K. Takasaki, T. Takebe, Rev. Math. Phys. 7 (1995) 743.

\bibitem{GD}
 I.M. Gelfand, L.A. Dickey, Func. Anal. Appl. 10 (1976) 4.

\bibitem{D}
L.A. Dickey, Soliton Equations and Hamiltonian Systems,
World Scientific, Singapore, 1991.

\bibitem{DIZ}
P. Di Francesco, C. Itzykson, J.B. Zuber, Commun. Math. Phys.
140 (1991) 543.

\bibitem{A}
M. Adler, Invent. Math. 50 (1979) 219.

\bibitem{KW}
B.A. Kupershmidt, G. Wilson, Invent. Math. 62 (1981) 403.

\bibitem{H}
W.J. Huang, J. Math. Phys. 35 (1994) 993.


















\end{thebibliography}
\end{document}